\documentstyle[preprint,aps]{revtex}
\begin{document}

\title{Ordered Products, $W_{\infty}$-Algebra, and
Two-Variable, Definite-Parity, Orthogonal Polynomials}
\author{A. Ver\c{c}in}

\address{Department of Physics \\
Ankara University, Faculty of Sciences,\\
06100, Tando\u gan-Ankara, Turkey\\
E.mail:vercin@science.ankara.edu.tr\\}
\maketitle

\begin{abstract}

It has been shown that the Cartan subalgebra of $W_{\infty}$-
algebra is the space of the two-variable, definite-parity
polynomials. Explicit expressions of these polynomials, and their
basic properties are presented. Also has been shown that they carry
the infinite dimensional irreducible representation of the $su(1,1)$
algebra having the spectrum bounded from below. A realization of
this algebra in terms of difference operators is also obtained.
For particular values of the ordering parameter $s$ they are
identified with the classical orthogonal polynomials of a
discrete variable, such as the Meixner, Meixner-Pollaczek, and
Askey-Wilson polynomials. With respect to variable $s$
they satisfy a second order eigenvalue equation of hypergeometric
type. Exact scattering states with zero energy for a family of
potentials are expressed in terms of these polynomials. It has been
put forward that it is the \.{I}n\"{o}n\"{u}-Wigner contraction and
its inverse that form bridge between the difference and differential
calculus.

\pacs{03.65. Ca, 02.20. Tw}

\end{abstract}

\section{INTRODUCTION}

The classical orthogonal polynomials of a discrete variable
(the Hahn, Meixner, Krawtchouk, Charlier, Pollaczek, and
Meixner-Pollaczek polynomials) which are the difference
analogues of the classical orthogonal polynomials of the
mathematical physics have been surfaced in various problems
of the theoretical and mathematical physics, in group
representation theory  and in computational physics and
techniques (\cite {Nikiforov,Askey,Broad}, and references
therein). Close relationships has been established between
the generalized spherical harmonics for su(2) and the
Krawtchouk polynomials, and between the Wigner 6j-symbols
and the Racah polynomials which are the discrete analogues
of the Jacobi polynomials on a quadratic lattice \cite
{Nikiforov,Yu}. The Clebsch-Gordan coefficients and 6j-symbols
for the su(1,1) can also be expressed in terms of the Hahn
and Racah polynomials \cite{Yu}. An important development in
connection with these polynomials in the last decade was that
the Hahn and Meixner polynomials can be analytically continued
in the complex plane, both in variable and parameter, such that
they become real polynomials which satisfy orthogonality relations
with respect to continuous measure \cite {Suslov,Askey}. These
polynomials are referred to as the continuous Hahn and Meixner, or, as
the Hahn and Meixner polynomials of an imaginary argument (in fact, the
second ones should be called the Meixner-Pollaczek polynomials \cite
{Askey}). The continuous Hahn polynomials are also closely related to
the unitary irreducible representations of the Lorentz group SO(3,1).

Recent interest in these polynomials are mainly due to
modifications and generalizations of them in connection
with their q-analogues on non-uniform lattices (\cite{Nodarse}
and references therein). Another line of development, which has not
attracted sufficient interest it deserves, was determination of their
connections with the ordered products of the Heisenberg-Weyl (HW)
algebra \cite{Bender1,Bender2}.
Connection between the ordered products of the HW-algebra
and particular classes of the continuous Hanh polynomials
was, for the first time, investigated in Ref.\cite{Bender1}
to find a solution for the unitarity problem in the finite
element approximation to quantum field equations. This
connection was established when ordering rules  that gives
Hermitian products are used \cite {Bender1,Bender2}.
In terms of ordering parameter $s\in {\bf C}$, henceforth used
in this paper, this corresponds to pure imaginary values of $s$
of which the well-known Weyl ordering is a special $s=0$ case.
One of the purposes of this study is to revive interest in
this direction by carrying out a systematic investigation in
the most general framework.

In two recent studies \cite {Vercin8,Vercin9}, we managed to
develop explicit expressions for the implicitly defined
conventional $s$-ordered products widely used in the
Weyl-Wigner-Groenevold-Moyal quantization and quantum optics
\cite{Vercin8,Vercin9,Cahill}. In section II, we briefly review
the ordered products, $W_{\infty}$-algebra and its Cartan
subalgebra in their  most general forms. Making use of these
expressions, we firstly associate with each element of the
Cartan subalgebra of the $W_{\infty}$-algebra in a general
$s$-basis  a two-variable, definite-parity, and (continuous
and discrete) orthogonal polynomials which are two-variable
generalization of those appeared in literature. Our variables
are $s$, and
the c-number correspondence of the operator $\hat{x}=
(\hat{q}\hat{p}+ \hat{p}\hat{q})/2$ denoted by $x$ (or, in some
cases the variable $u=\frac{x}{i\hbar}-\frac{1}{2}$). These
polynomials solve a hypergeometric type differential equation
with respect to $s$, and a hypergeometric type difference equation
with respect to $x$. For generic values of $s$ this difference
equation is identical with that satisfied by a particular class
of Meixner polynomials. Other basic properties, and difference-
differential relations are also presented. We also show that they
carry the irreducible representation of the su(1,1)-algebra
and obtain its realization in terms of difference operators
(section III). Secondly, for particular values of $s$, we identify
these polynomials with particular classes of the Meixner, Meixner-
Pollaczek, Askey-Wilson, and Hahn polynomials, and give the
discrete and continuous orthogonality relations for them
(section IV). Thirdly, we show that exact zero-energy scattering
states for a family of the P\"{o}sch-Teller type potentials are
expressed in terms of these polynomials (section V). Finally, we
conclude by pointing out the role played by the \.{I}n\"{o}n\"{u}-
Wigner contraction and its inverse transformation in connection
with differential-difference calculus.

\section{Ordered Products, $W_{\infty}-$Algebra and its Cartan Subalgebra}

Let us consider the HW-algebra: $[\hat{q},\hat{p}]=i\hbar\hat{I}$,
where $\hbar, \hat{I}, \hat{q}$ and $\hat{p}$ are the
Planck's constant, the identity operator and the
Hermitian position and momentum operators, respectively.
Here and henceforth operators and functions of operators
acting in a Hilbert space $\cal {H}$ are denoted by a
$\hat{}$ over letters. We define the s-ordered products
$\hat{t}^{(s)}_{nm}\equiv\{(\hat{q})^{n}(\hat{p})^{m}\}_{s}$,
containing $n$ factors of $\hat{q}$ and $m$ factors of $\hat{p}$,
by the following explicit equivalent formulas
\begin{eqnarray}
\hat{t}^{(s)}_{nm}&=&2^{-n}\sum^{n}_{j=0}(^{n}_{j})(1+s)^{j}(1-s)^{n-j}
\hat{q}^{j}\hat{p}^{m}\hat{q}^{n-j}\nonumber\\
&=&2^{-m}\sum^{m}_{k=0}(^{m}_{k})(1-s)^{k}(1+s)^{m-k}
\hat{p}^{k}\hat{q}^{n}\hat{p}^{m-k}.
\end{eqnarray}
These give for $s=\pm1, \hat{t}^{(1)}_{nm}=\hat{q}^{n}\hat{p}^{m};$
$\hat{t}^{(-1)}_{nm}=\hat{p}^{m}\hat{q}^{n}$ and for $s=0$,
$\hat{t}^{(0)}_{nm}=2^{-n}\sum^{n}_{j=0}(^{n}_{j})\hat{q}^{j}\hat{p}^{m}\hat{q}^{n-j}$
$=2^{-m}\sum^{m}_{k=0}(^{m}_{k})\hat{p}^{k}\hat{q}$
$^{n}\hat{p}^{m-k}$. While the first two of these expressions exhibit
the standart and antistandart rules of ordering, respectively, that
corresponding to $s=0$ are two well-known expressions of the
Weyl, or symmetricaly ordered products. It is possible to write
many equivalent forms of the above relations, but, for later use
only two of them have been written. Although, there
are not any known physical applications apart from the three
principle ones corresponding to $s=1,0,-1,$ embedding orderings
in a continuum provides a natural context for viewing their
differences and interrelationships in a continuous manner and
enable us to carry out the related analyses in their most general
forms.

An arbitrary s-ordered product can be expressed in terms of a
polynomial in $s^{\prime}$-ordered product as follows
\begin{eqnarray}
\hat{t}^{(s)}_{nm}=
\sum^{(n,m)}_{k=0}2^{-k}(^{n}_{k})(^{m}_{k})k!
[i\hbar(s-s^{\prime})]^{k}\hat{t}^{(s^{\prime})}_{n-k,m-k},
\end{eqnarray}
where $(n,m)$ denotes the smaller of the integers $n$ and $m$,
$(^{n}_{k})=n![(n-k)! k!]^{-1}$ is a binomial coefficient, and
$s^{\prime}$ is also arbitrary complex number .
Note that $i\hbar$ in Eq. (2) is the sign of the commutator of the
corresponding operators there. Thus, the relations (1) and (2)
can be used for any pair of the operators $\hat{A},\hat{B}$ of any
algebra satisfying the commutation relation $[\hat{A},\hat{B}]=
i\lambda,\lambda \in \bf{C}$. From (1)  easily follows that
\begin{eqnarray}
[\hat{t}^{(s)}_{nm}]^{\dagger}=\hat{t}_{nm}^{(-\bar{s})},
\end{eqnarray}
where $\dagger$ stands for the Hermitian conjugation and $\bar{a}$
is the complex conjugation of $a$. That is, for general $n,m$
integers, $\hat{t}^{(s)}_{nm}$ are Hermitian if and only if
$\bar{s}=-s$. In particular, the Weyl ordered products
$\hat{t}^{(0)}_{nm}$ are Hermitian. For general $s,\alpha\in\bf {C}$
one can form combinations such as $\hat{\kappa}_{nm}(s)
=\alpha \hat{t}_{nm}^{(s)}+\bar{\alpha}\hat{t}_{nm}^{(-\bar{s})}$
that are Hermitian.

The $W_{\infty}$-algebra is the infinite algebra
generated by the ordered products $\hat{t}^{(s)}_{nm}$
\cite{Vercin9,Pope,Cappelli}. Up to a trivial central element it
is the universal enveloping algebra of the HW-algebra. In the most
general basis the structure constants of the $W_{\infty}$-algebra
can be read off from
\begin{eqnarray}
[\hat{t}^{(s)}_{kl}, \hat{t}^{(s)}_{nm}]=
-\sum^{j_{max}}_{j=0}\frac{i^{j}}{j!}[\sum^{j \prime}_{r=0}(^{j}_{r})
f_{srj}a_{nmkl,rj}]\hat{t}^{(s)}_{n+k-j,m+l-j},
\end{eqnarray}
where the prime over the second summation indicates that the
maximum value that $r$ may take is $r_{max}=(m,k)$ and
\begin{eqnarray}
j_{max}=(n+r_{max},l+r_{max}) , \qquad
a_{nmkl,rj}=\frac{n!m!k!l!}{(n+r-j)!(m-r)!(k-r)!(l+r-j)!}.
\end{eqnarray}
The restrictions imposed on summations also follows
from the expression of $a_{nmkl,rj}$. In relation (4)
$f_{srj}=(s^{-})^{r}(-s^{+})^{j-r}-(s^{-})^{j-r}(-s^{+})^{r}$,
$s^{\pm}=\hbar(1\pm s)/2$,
is the only factor depending on the  chosen rule of ordering.
Here we observe that anti-commutator of the ordered products
is given by the same relation as (4) only provided that the
ordering factor is replaced by
$f^{+}_{srj}=(s^{-})^{r}(-s^{+})^{j-r}+(s^{-})^{j-r}(-s^{+})^{r}$.
Relation (4), which was first reported in Ref.\cite{Vercin9}, for
$s=0,\pm1$ coincides with those appeared in literature \cite
{Dunne}, and generalize them for arbitrary values of $s$.

$W_{\infty}$ has some finite and infinite dimensional
subalgebras. But, for the purpose of this work we will be
concerned only with the infinite abelian subalgebra
consisting of the generators $\hat{H}^{(s)}_{n}
\equiv \hat{t}^{(s)}_{nn}$. The commutativity of the generators
$[\hat{H}^{(s)}_{n},\hat{H}^{(s)}_{k}]=0$ follows from (4).
But, for our purpose we prove this, and
$[\hat{H}^{(s)}_{n},\hat{H}^{(s^{\prime})}_{k}]=0$
by showing that for any values
of $n \geq 0$ and $s \in {\bf C}$ all the ordered products of the
form $\hat{H}^{(s)}_{n}\equiv \hat{t}^{(s)}_{nn}$ can be expressed
in terms of single operator
$\hat{x}\equiv (\hat{q}\hat{p}+\hat{p}\hat{q})/2$.
To see this, let us first consider
$\hat{H}^{(1)}_{n}=\hat{q}^{n}\hat{p}^{n}$
which can be written as
\begin{eqnarray}
\hat{H}^{(1)}_{n}&=&\hat{q}^{n-1}\hat{q}\hat{p}\hat{p}^{n-1}
=\hat{q}^{n-1}\hat{p}^{n-1}[\hat{x}+\hat{c}(2n-1)] \nonumber\\
&=&\hat{q}^{n-2}\hat{p}^{n-2}[\hat{x}+\hat{c}(2n-3)]
[\hat{x}+\hat{c}(2n-1)].\nonumber
\end{eqnarray}
Hence, by induction, we have
\begin{eqnarray}
\hat{H}^{(1)}_{n}=\prod^{n}_{j=1}[\hat{x}+\hat{c}(2j-1)],\qquad
\hat{H}^{(-1)}_{n}=\prod^{n}_{j=1}[\hat{x}-\hat{c}(2j-1)],
\end{eqnarray}
where $\hat{c}\equiv i\hbar\hat{I}/2$  and the relations
$\hat{q}\hat{p}=\hat{x}+\hat{c}$,
$[\hat{x},\hat{p}^{k}]=2\hat{c}k\hat{p}^{k}$ are used. The second
relation is written by making use of (3).

\section{Two-Variable, Definite Parity Polynomials}

Now, evaluating (2) for $s^{\prime}=\pm 1$ by making use of (6), and then
replacing $\hat{x}$ in  results by the c-number variable $x$, we
obtain  two-variable polynomials  $P_{n}(s,x)$ for each element
of the Cartan subalgebra. Here the variables are the
(dimensionless) ordering parameter $s\in {\bf C}$ and $x\in {\bf R}$
which has dimension of angular momentum (since $\hat{x}$ is an
hermitian operator we consider $x$ as a real variable). Two
equivalent, explicit expressions for these polynomials are as follows
\begin{eqnarray}
P_{n}(s,x)&=&\sum^{n}_{k=0}
(^{n}_{k})^{2}k![-c(1-s)]^{k}\prod^{n-k}_{j=1}[x+c(2j-1)]\nonumber\\
&=&\sum^{n}_{k=0}(^{n}_{k})^{2}k![c(1+s)]^{k}
\prod^{n-k}_{j=1}[x-c(2j-1)],
\end{eqnarray}
where $c=i\hbar/2$. From these relations it is
obvious that under the action of two
dimensional parity transformation in ${\bf CxR}$, they transform as
\begin{eqnarray}
P_{n}(-s,-x)=(-1)^{n}P_{n}(s,x),
\end{eqnarray}
that is, they have the same parity with $n$.
The following recursion relation can also be verified
\begin{eqnarray}
P_{n+1}(s,x)=[x+c(2n+1)s]P_{n}(s,x)+c^{2}(1-s^{2})n^{2}P_{n-1}(s,x),
\end{eqnarray}
The easiest way of obtaining this relation may be first noting the
relation
\begin{eqnarray}
[\hat{x}, \hat{t}^{(s)}_{nm}]_{+}=2[\hat{t}^{(s)}_{n+1,m+1}-
cs(m+n+1)\hat{t}^{(s)}_{nm}-c^{2}nm(1-s^{2})\hat{t}^{(s)}_{n-1,m-1}],\nonumber
\end{eqnarray}
where $[, ]_{+}$ stands for the anticommutator. They also obey the
following derivatives and difference relations
\begin{eqnarray}
\partial^{k}_{s}P_{n}(s,x)&=&c^{k}[\frac{n!}{(n-k)!}]^{2}P_{n-k}(s,x),\\
(x \pm c)P_{n}(s, x \pm 2c) &=& [x \pm c(2n+1)]P_{n}(s,x) +
2c^{2}n^{2}(1 \mp s)P_{n-1}(s,x).
\end{eqnarray}
with respect to $s$ and $x$, respectively.
This last property immediately results by first noting the relations
\begin{eqnarray}
(x \pm c)P_{n}(\pm1,x \pm 2c) =[x \pm c(2n+1)]P_{n}(\pm1,x),
\end{eqnarray}
which easily result from (6).

From (9) and (10) we see that, with respect to $s$ the lowest
order differential equation obeyed by these polynomials is the
following hypergeometric type
differential equation
\begin{eqnarray}
\{(1-s^{2})\partial^{2}_{s}+
[\frac{x}{c}+(2n-1)s]\partial_{s} -n^{2}\}P_{n}(s,x)=0.
\end{eqnarray}
The difference relations (12) can also be recast in a form in which
the discrete differences are in a more conventional form by using
the central first and second differences;
\begin{eqnarray}
D_{h}f(x)=\frac{f(x+h)-f(x-h)}{2h},\qquad
D^{2}_{h}f(x)=\frac{f(x+h) -2f(x)+f(x-h)}{h^{2}},
\end{eqnarray}
which approximate the usual first and the second derivatives
on a lattice with the constant mesh $\Delta x = h$ up to second
order in $h$ \cite {Nikiforov,Bender2}. For this purpose we take
the sum and difference of two relations given by (12)
\begin{eqnarray}
(x+c)P_{n}(s,x+2c)-(x-c)P_{n}(s,x-2c) &=& c(4n+2)P_{n}(s,x) -
4sc^{2}n^{2}P_{n-1}(s,x),\\
(x+c)P_{n}(s,x+2c)+(x-c)P_{n}(s,x-2c) &=& 2x P_{n}(s,x)+
4c^{2}n^{2}P_{n-1}(s,x),
\end{eqnarray}
which, in terms of $D_{h}$ and $D^{2}_{h}$ can be
rewritten as follows
\begin{eqnarray}
(c^{2}D^{2}_{h}+xD_{h} - n)P_{n}(s,x)&=&-scn^{2}P_{n-1}(s,x),\\
(xD^{2}_{h}+D_{h})P_{n}(s,x) &=&n^{2}P_{n-1}(s,x),
\end{eqnarray}
where, $h=2c$ and $D_{h}$ denotes the partial difference
operation with respect to $x$. Note that all the difference
operations are with respect to variable $x$ and all the
derivatives are with respect to the ordering variable $s$.

Now, by making use of the recursion relation (9), and
(17), (18) we have
\begin{eqnarray}
J_{-}P_{n}(s,x)&=&n^{2}P_{n-1}(s,x),\nonumber\\
J_{+}P_{n}(s,x)&=&P_{n+1}(s,x),\nonumber\\
J_{0}P_{n}(s,x)&=&(n+\frac{1}{2})P_{n}(s,x),
\end{eqnarray}
where
\begin{eqnarray}
J_{-}&=&[xD_{h}^{2}+D_{h}]
=[\frac{4x}{\hbar^{2}}
\sin^{2}(\frac{\hbar}{2}\partial_{x})+
\frac{1}{\hbar}\sin(\hbar \partial_{x})],\\
J_{+}&=&c^{2}(1+s^{2})J_{-}+2cs(c^{2}D_{h}^{2}+
xD_{h})+(x+cs)\nonumber\\
&=&c^{2}(1+s^{2})J_{-}+2cs
[-\sin^{2}(\frac{\hbar}{2}\partial_{x})+
\frac{x}{\hbar}\sin(\hbar\partial_{x})]+(x+cs),\\
J_{0}&=&c(c+sx)D_{h}^{2}+(x+cs)D_{h}+\frac{1}{2} \nonumber\\
&=&c(c+sx)\frac{4}{\hbar^{2}}\sin^{2}(\frac{\hbar}{2}\partial_{x})+
(x+cs)\frac{1}{\hbar}\sin(\hbar\partial_{x})+\frac{1}{2}.
\end{eqnarray}
It is not hard to verify that by their actions on an arbitrary
function, the generators $(J_{0},J_{\pm})$ obey the standart
defining relations of the $su(1,1)$ algebra
\begin{eqnarray}
[J_{+}, J_{-}]=-2J_{0},\qquad [J_{0}, J_{\pm}]= \pm J_{\pm}.
\end{eqnarray}
The value of the Casimir operator
$J^{2}=-J_{-}J_{+}+J_{0}^{2}+J_{0}$
is found to be
$J^{2}P_{n}(s,x)=-\frac{1}{4}P_{n}(s,x)$.
Thus, we have obtained a realization of the $su(1,1)$ algebra in
terms of the difference operators, and the polynomials we have
found carry the infinite dimensional irreducible representation of
this algebra having the spectrum bounded from below which is
designated by $D^{+}_{-\frac{1}{2}}$ \cite {Wybourne}.

Finally, in this section we write down explicit expressions of the
first five polynomials;
\begin{eqnarray}
P_{0}(s,x)&=&1,\nonumber\\
P_{1}(s,x)&=&x+cs, \nonumber\\
P_{2}(s,x)&=&x^{2}+4csx+c^{2}(2s^{2}+1), \nonumber\\
P_{3}(s,x)&=&x^{3}+9csx^{2}+c^{2}(18s^{2}+5)x+3c^{3}s(2s^{2}+3), \nonumber\\
P_{4}(s,x)&=&x^{4}+16csx^{3}+2c^{2}(36s^{2}+7)x^{2}+16c^{3}s(6s^{2}+5)x+
3c^{4}(8s^{4}+24s^{2}+3).
\end{eqnarray}
The values of these polynomials at $(s,x=0)$ and at $(s=0,x=0)$
can be easily obtained from (7) \cite{Note1,Note2}. Because of (3), these
are real valued for pure imaginary $s$. Note that the relation
(2), which enables us to write $P_{n}(s,x)$ in terms of
$P_{n}(s^{\prime},x)$, in view of (10), is nothing more than a
Taylor expansion.

\section{Connections with the Meixner, Meixner-Pollaczek
and Continuous Hahn Polynomials}

To compare the polynomials we have found with the classical
orthogonal polynomials of a discrete  variable we use
\begin{eqnarray}
u=\frac{x}{2c}-\frac{1}{2}.
\end{eqnarray}
In that case we have the following difference relations
\begin{eqnarray}
(u+1)P_{n}(s,u+1)&=&(u+n+1)P_{n}(s,u) +
c(1-s)n^{2}P_{n-1}(s,u),\\
uP_{n}(s,u-1)&=&(u-n)P_{n}(s,u)+
c(1+s)n^{2}P_{n-1}(s,u).
\end{eqnarray}
Eliminating the last terms between these two relations
and writing out the result in terms of the difference operators
\begin{eqnarray}
\Delta f(u)=f(u+1)-f(u),\qquad \nabla f(u)=f(u)-f(u-1),
\end{eqnarray}
we arrive at
\begin{eqnarray}
\{u\Delta \nabla-[\frac{2}{1-s}u+\frac{1+s}{1-s}]\Delta+
\frac{2n}{1-s} \}P_{n}(s,u)=0.
\end{eqnarray}
This is the same type of difference equation satisfied by the
Meixner polynomials $m_{n}^{(\gamma,\mu)}(u)$:
\begin{eqnarray}
\{u \Delta \nabla +[(\mu-1)u+\mu \gamma]\Delta+\lambda_{n} \} m_{n}^{(\gamma,\mu)}(u)=0.
\end{eqnarray}
Comparing with Eq. (29) we have
\begin{eqnarray}
\mu=\frac{s+1}{s-1},\qquad \gamma=1,\qquad\lambda_{n}= \frac{2n}{1-s}.
\end{eqnarray}

The Meixner polynomials are normalized as follows
$\Delta^{n}m_{n}^{(\gamma,\mu)}(u)=n!(\frac{\mu-1}{\mu})^{n}$.
On the other hand, the polynomials $P_{n}(s,u)$ obey the relation
$\Delta^{n}P_{n}(s,u)=(2c)^{n}n!$. (For a polynomial $q_{m}(u)$ of
degree $m$ the $\it{m}$th diferences and $\it{m}$th derivatives are
equal: $\Delta^{m}q_{m}(u)=\nabla ^{m}q_{m}(u)= \partial_{u}^{m}q_{m}(u)$).
Thus, we obtain
\begin{eqnarray}
P_{n}(s,u)=[c(s+1)]^{n}m_{n}^{(1,\frac{s+1}{s-1})}(u).
\end{eqnarray}
If the conditions $\gamma > 0$, and $0< \mu <1$ are satisfied the
Meixner's polynomials obey the discrete-orthogonality relation
\begin{eqnarray}
\sum_{u=0}^{\infty}m_{n}^{(\gamma,\mu)}(u)m_{m}^{(\gamma,\mu)}(u)\rho (u)=
\delta_{nm}d_{n}^{2}.
\end{eqnarray}
Thus, provided that $s<-1$ this relation in terms of $P_{n}(s,u)$ is
as follows
\begin{eqnarray}
\sum_{u=0}^{\infty}P_{n}(s,u)P_{m}(s,u)\rho^{\prime}(u)=
\delta_{nm}d_{n}^{\prime 2},
\end{eqnarray}
where, (for $s<-1$ ) the weight $\rho^{\prime}(u)$, and the squared
norm $d_{n}^{\prime 2}$ are found to be
\begin{eqnarray}
\rho^{\prime}(u)=(\frac{s+1}{s-1})^{u}\frac{\Gamma(u+1)}{u!},\qquad
d_{n}^{\prime 2}=
\frac{1}{2}(n!)^{2}c^{2n}(1-s)(s^{2}-1)^{n},
\end{eqnarray}
Here $\Gamma(z)$ is the gamma function.
It is also easy to verify that, the three-term recurrsion
relation (9) for the polynomials $P_{n}(s,u)$ is the same as
that obeyed by the Meixner's polynomials
\begin{eqnarray}
\mu m_{n+1}^{(\gamma,\mu)}(u)=
[\gamma \mu+(1+\mu)n-(1-\mu)u]m_{n}^{(\gamma,\mu)}(u)-
n(n+\gamma-1)m_{n-1}^{(\gamma,\mu)}(u),
\end{eqnarray}
provided that relations given by (31) are satisfied.

In Ref. \cite {Suslov} by considering the analytic continuation
of the orthogonality relation (33) in the parameter $\mu=\exp(-2i\phi)$
the polynomials
\begin{eqnarray}
P_{n}^{\lambda}(\phi,t)=
\frac{e^{-in\phi}}{n!}m_{n}^{(2\lambda,\mu)}(-\lambda+it),
\end{eqnarray}
which, under the conditions $\lambda>0$, and $0<\phi<\pi$, obey
the orthogonality relation
\begin{eqnarray}
\int_{-\infty}^{\infty}P_{n}^{\lambda}(\phi,t)P_{m}^{\lambda}(\phi,t)\rho_{P}(t)dt
=\delta_{nm}\frac{\Gamma(2\lambda+n)}{n!}
\end{eqnarray}
with respect to continuous measure are obtained. Here the weight
$\rho_{P}$ is as follows
\begin{eqnarray}
\rho_{P}(t)=\frac{1}{2 \pi}(2\sin \phi)^{2\lambda}|\Gamma(\lambda+it)|^{2}
\exp[(2\phi-\pi)t].
\end{eqnarray}
These polynomials should be called the Meixner-Pollaczek polynomials
\cite {Askey}. In view of the relation (32), if we identify
$\hat{x}/\hbar$ with the real variable $t$ we find that $\lambda=1/2$
and $s=i\cot\phi$. Thus, with help of (37), we obtain
\begin{eqnarray}
P_{n}(s,i\frac{x}{\hbar}-\frac{1}{2})=
n!(\frac{-\hbar}{2\sin \phi})^{n}P_{n}^{1/2}(\phi,\frac{x}{\hbar}).
\end{eqnarray}
In terms of $P_{n}(s,u)$ the orthogonality relation (38) is
as follows
\begin{eqnarray}
\int_{-\infty}^{\infty}P_{n}(s,i\frac{x}{\hbar}-\frac{1}{2})
P_{m}(s,i\frac{x}{\hbar}-\frac{1}{2})\rho_{0}(x)dx=
\delta_{nm}(n!)^{2}(\frac{\hbar}{2\sin \phi})^{2n+1},
\end{eqnarray}
with the weight
\begin{eqnarray}
\rho_{0}(x)=\frac{\exp[(2\phi-\pi)\frac{x}{\hbar}]}{\cosh \frac{\pi x}{\hbar}}.
\end{eqnarray}

For $\lambda=1/2$, and $\phi=\pi/2$ the Meixner-Pollaczek polynomials
are related to the Askey-Wilson polynomials $q_{n}^{(\alpha)}(x,\delta)$
as follows $q_{n}^{(0)}(x,1/2)=(1/2)_{n}P_{n}^{1/2}(\pi/2, x)$,
where, $(a)_{n}=\Gamma(a+n)/\Gamma(a)$. Note that this particular
case corresponds to $s=0$ (Weyl ordering) and the weight is
$\rho_{0}(x)=\cosh(\pi x/\hbar)$. On the other hand, the Askey-Wilson
polynomials are particular cases of the continuous Hahn polynomials
$h_{n}^{(\alpha,\beta)}(z,N)$:
\begin{eqnarray}
q_{n}^{(\alpha)}(x,\delta)=(-i)^{n}h_{n}^{(\alpha,\alpha)}
(\frac{1}{2}ix-\frac{1}{2}(\alpha-\delta+1),-\alpha+\delta).
\end{eqnarray}
Making use of these relations we have
\begin{eqnarray}
P_{n}(s=0,i\frac{x}{\hbar}-\frac{1}{2})&=&
n!(\frac{-\hbar}{2})^{n}[(\frac{1}{2})_{n}]^{-1}q_{n}^{(0)}(\frac{x}{\hbar},\frac{1}{2})\nonumber\\
&=& n!(\frac{-\hbar}{2})^{n}[(\frac{1}{2})_{n}]^{-1}(-i)^{n}h_{n}^{(0,0)}
(i\frac{x}{2\hbar}-\frac{1}{4},\frac{1}{2}).
\end{eqnarray}

As a last remark in this section we note that for particular
values, or ranges of the variables, these polynomials can also
be identifed with particular cases of the Jacobi polynomials
and the generalized spherical harmonics (the Bargmann
functions)\cite{Broad}.

\section{Rodrigues Formula, Generating Function, and
Exact Zero-Energy Scattering States For a Family of Potentials}

For a given $x$, the polynomials solve the eigenvalue
equation (13), which is a differential equation of the
hypergeometric type. In order to uncover more
properties of these polynomials we restrict the
investigation to
\begin{eqnarray}
s=iy,\qquad \frac{x}{c}=-iv;\qquad y,v \in{\bf R}
\end{eqnarray}
Note that in that case the corresponding ordered products
are Hermitian. In terms of the new variables, Eq. (13) is as
follows
\begin{eqnarray}
\{(1+y^{2})\partial^{2}_{y} + [v-(2n-1)y]\partial_{y}+n^{2} \}P_{n}(y,v)=0
\end{eqnarray}
With the help of the function
\begin{eqnarray}
\rho(y,v)=\frac{e^{vtan^{-1}y}}{(1+y^{2})^{\frac{2n+1}{2}}},
\end{eqnarray}
which is determined from
\begin{eqnarray}
\partial_{y}[(1+y^{2})\rho]=[v-(2n-1)y]\rho,
\end{eqnarray}
Eq. (46) can be written in the self-adjoint form
\begin{eqnarray}
\{\partial_{y}[(1+y^{2})\rho(y,v)\partial_{y}] + \rho(y,v)n^{2} \}P_{n}(y,v)=0.
\end{eqnarray}
Thus, we have found the following Rodrigues formula, integral
representation, and the generating function for these polynomials:
\begin{eqnarray}
P_{n}(y,v)&=&\frac{(-ic)^{n}}{\rho(y,v)}
\partial^{n}_{y}[\frac{e^{vtan^{-1}y}}{(1+y^{2})^{\frac{1}{2}}}],\\
P_{n}(y,v)&=&\frac{(-ic)^{n} n!}{2 \pi i \rho(y,v)}\oint_{C}
\frac{(1+z^{2})^{n}}{(z-y)^{n+1}}\rho(z,v)dz,\\
\Phi(y,v,u)&=&(1-4uy-4u^{2})^{-1/2}\frac{\rho(\xi,v)}{\rho(y,v)} ,\\
(\xi &=&(2u)^{-1}[1-(1-4uy-4u^{2})^{1/2}]) .\nonumber
\end{eqnarray}
In Eq. (51) $C$ is a closed contour surrounding the point $z=y$, and
the expansion of the generating function $\Phi(y,x,u)$
in power of $u$ has, for sufficiently small $|u|$, the form
\begin{eqnarray}
\Phi(y,v,u)=\sum^{\infty}_{n=0}\frac{P_{n}(y,v)}{n!}(-u/ic)^{n}.
\end{eqnarray}

Finally in this section, we would like to interpret these
polynomials as the solutions of the Scr\"{o}dinger type equations.
By transforming the dependent variable as
\begin{eqnarray}
\Psi_{n}(y,v)=[(1+y^{2})\rho(y,v)]^{1/2}P_{n}(y,v),
\end{eqnarray}
the first derivative term in Eq. (46) vanishes and the equation
\begin{eqnarray}
\partial^{2}_{y}\Psi_{n}(y,v) + \frac{y^{2}+2v(2n+1)y +
(2n+1)^{2}-(v^{2}+3)}{4(1+y^{2})^{2}}\Psi_{n}(y,v)=0,
\end{eqnarray}
results. $\eta$ being a constant of dimension
$(length)^{-1}$, Eq. (55) is the time independent Schr\"{o}dinger
equation for the potential
\begin{eqnarray}
\frac{2m}{\eta^{2}\hbar^{2}}V_{n}(y,v) =-\frac{y^{2}+2v(2n+1)y +
(2n+1)^{2}-(v^{2}+3)}{4(1+y^{2})^{2}}.
\end{eqnarray}
In fact, here we have a family of potentials which are
labelled by $n$ and $v$, and are similar to the P\"{o}sch-Teller
type ($\propto -\cosh^{-2}y$) potential holes . Thus, the
wave functions given by (54) are the exact zero-energy
scattering states for this family of potentials. Note
that for $v=0$, $V(y,0)$ is an even function, so admit the
solutions $\Psi_{n}(y,0)$ with the same parity as $n$.

\section{conclusion}

The main points of this study can be summarized as follows.
(i) By using the explicit expression for the ordered
products, which form a basis for the universal enveloping
algebra of the HW-algebra, we have shown that, the infinite
Cartan subalgebra is, in fact, the abelian algebra of two-
variable polynomials. The variables are the ordering
parameter $s$ and the c-number correspondence of the squeeze
operator $\hat{x}$ which is an element of the symplectic
algebra in two dimensions. We expect that these definite-parity
polynomials will play an important role in two
dimensional physics in which tremendous developments are
taking place in the recent years (see, for instance
\cite{Cappelli}, and references therein).
(ii) The realization of the $su(1,1)$ obtained in this
report, is, since it contains a single variable, a difference
analoque of generalized Gelfand-Dyson realization
\cite{Alhassid}. Furthermore, the analyses of section V  show
that this algebra play the role of ``potantial algebra" (i.e., an
algebra whose generators connect the states of the same energy in
different potential strengths \cite{Alhassid}) for a class of
potentials. Regarding the role of the $su(1,1)$ algebra in the
exactly solvable problems of quantum mechanics, we
expect that with its realization in terms of the difference
operators found in this study, or, with appropriate
generalizations, it is also the underlying
algebra of the exactly solvable difference equations. This
point, of course, requires futher studies to be done. In
particular, an algebraic approach to the polynomials of a
discrete variable is now under investigation.
(iii) Another particularly intriguing point requiring
further studies is that
the basic algebraic bulding block underlying the difference
equations seems to be an expansion of the HW-algebra, which is
the basic algebraic structure of both classical and quantum
mechanics. More conceretly, the algebra:
\begin{eqnarray}
[x,D_{h}]=-\Delta_{h},\qquad [x,\Delta_{h}]=\hbar^{2}D_{h},\qquad [D_{h},
\Delta_{h}]=0 ,\nonumber
\end{eqnarray}
satisfied by the fundamental difference operations
$\{x, D_{h}, 2\Delta_{h}f(x)=f(x+h)+f(x-h) \}$
is the two-dimensional Euclidean algebra $e(2)=\{L, N_{1}, N_{2}:$
$[L, N_{1}]=N_{2}, [L, N_{2}]=-N_{1}, [N_{1},N_{2}]=0 \}$
subjected to singular transformation (with respect to $N_{2}$):
$\{x=\hbar L, \Delta_{h}=\hbar N_{1}, D_{h}=N_{2}\}$. The
$\hbar \rightarrow 0$ (\.{I}n\"{o}n\"{u}-Wigner) contraction
of this algebra is the HW-algebra generated by
\begin{eqnarray}
x,\qquad \partial_{x}=\lim_{\hbar \rightarrow 0}D_{h},\qquad
I=\lim_{\hbar \rightarrow 0}\Delta_{h}\nonumber.
\end{eqnarray}
Thus, we expect that it
is the \.{I}n\"{o}n\"{u}-Wigner and its inverse transformation which
form a bridge between the exactly solvable differential and
difference equations of mathematical physics. Other results
in this direction and more detailed properties of these polynomial
will be given elsewhere.

\acknowledgments

I wish to thank T. Dereli, M. \"{O}nder, and C. Harabati for helpful
discussions. This work was supported in part by the Scientific
and Technical Research Council of Turkey (T\"{U}B\.{I}TAK).

\end{document}